\documentclass[ aip,
 amsmath,amssymb,
 reprint,%
]{revtex4-1}

\usepackage{soul}

\usepackage[dvipdfmx]{graphicx}
\usepackage{dcolumn}
\usepackage{bm}
\usepackage{color}
\definecolor{Red}  {rgb}{1,0,0}
\definecolor{Green}{rgb}{0,1,0}
\definecolor{Blue} {rgb}{0,0,1}
\newcommand {\delspan}[1] {}

\newcommand {\bfv}[1] {{\boldsymbol {#1}}}

\newcommand {\ds} {\displaystyle}

\newcommand {\IND}[1] {\ }
\newcommand\Rey{\mbox{{\rm Re}}}     
\newcommand\Grs{\mbox{{\rm Gr}}}     
\newcommand\Prt{\mbox{{\rm Pr}}}  
\newcommand\removed[1]{}

 \newcommand\SEC[1]{\section{ #1}}

\usepackage{ulem}

\usepackage{dcolumn}
\usepackage{bm}
\usepackage[utf8]{inputenc}
\usepackage[T1]{fontenc}
\usepackage{mathptmx}
\usepackage{etoolbox}
\makeatletter
\def\@email#1#2{%
 \endgroup
 \patchcmd{\titleblock@produce}
  {\frontmatter@RRAPformat}
  {\frontmatter@RRAPformat{\produce@RRAP{*#1\href{mailto:#2}{#2}}}\frontmatter@RRAPformat}
  {}{}
}%
\makeatother
\begin{document}

\preprint{AIP/123-QED}

\title{A thermal convection limit of spiral state in wide-gap spherical Couette flow}



\author{Tomoaki Itano}
\email{itano@kansai-u.ac.jp}
\author{Fumitoshi Goto}
\author{Kazuki Yoshikawa}
\affiliation{
  Department of Pure and Applied Physics,
  Faculty of Engineering Science, Kansai University,  Osaka, 564-8680, Japan
}
\author{Masako Sugihara-Seki}
\altaffiliation[Also at]{
  Graduate School of Engineering Science, Osaka University, Osaka 560-8531, Japan
}
\affiliation{
  Department of Pure and Applied Physics,
  Faculty of Engineering Science, Kansai University,  Osaka, 564-8680, Japan
}  


\date{\today}

\begin{abstract}
The symmetries of flow structures are often prescribed by their mechanical instability and geometry.
Here, as an example, we present  the homotopy of the rotating 3-fold spiral state that is robust in a spherical Couette flow towards the hybrid system with a thermal stratification effect.
It has not yet been confirmed that the rotating wave state smoothly connects to the thermal stratification system.
  Through continuation, the most dangerous mode at a purely spherical Couette flow of $m=4$ modes of spherical harmonics is replaced by $l=4$ and $m=3$ in a purely thermal convective system.
  For the state obtained at the limit under only the thermal effect,  the residual quantities of both the torque to the outer sphere and meridional circulation are discussed in detail.
\end{abstract}



\maketitle

%
%
%
%

\SEC{Introduction}
\IND{}
Fluid motion subject to multiple diffusion processes on a sphere exhibits rich spatiotemporal properties despite their geometrical restrictions.
The mathematical similarity underlying their independent mechanisms has attracted significant attention from researchers because of the importance of resolving the formation of inhomogeneous structures in biological organisms\cite{Ala52} and spatial patterns in geophysical and astrophysical bodies\cite{Fow04}.
Conventionally studied thermal convective systems, which are configurations that are applicable in astrophysical and geophysical flows, may also be categorized into the same class  in which two conservative quantities, momentum and heat that function as suppression and activation factors, competitively diffuse.

\IND{}
Thermal convection in a spherical geometry has been studied with the addition of other mechanical or chemical effects\cite{Cha61}.
For instance, the influence of the radial-dependence proportionality of the gravity field on convection has been studied experimentally and numerically under microgravity conditions at the International Space Station\cite{Feu11}.
The addition of both electric conductivity and system rotation to the spherical thermal convection system may generate pairs of cyclonic and anticyclonic vortex columns in the shell\cite{Kid97}.
They can reproduce the self-generation and reversal of magnetic fields related to geophysical phenomena\cite{Sak99}.
According to the Taylor-Proudman theorem, vortex columns induced by thermal convection under the rotation of the entire system align with the rotation axis via the Coriolis effect, which would be realizable even in the absence of an electromagnetic effect\cite{Kim11}.
From the viewpoint of geophysics, it is interesting that the generated vortex columns rotate either faster or slower than the system \cite{Feu13,Feu15}.
In these examples, the transition is initiated by vortices satisfying the mirror symmetry on the equatorial plane because the Coriolis effect suppresses the freedom of the system in the polar direction.

\IND{}
However, the flow that is confined between spherical boundaries that rotate differentially, namely, {\it spherical Couette flow} (SCF)\cite{Egb95,Wul99,Nak02,Abb18a}, is of interest because it can reproduce the turbulent state even without the thermal effect.
In particular, the case between a rotating inner sphere and a stationary outer sphere has been extensively investigated.
Past investigations with several different combinations of spherical boundaries with various radii\cite{Mun75,Bel84,Egb95} have shown that SCFs with a relatively wide gap are different from cylindrical Taylor-Couette flows despite the apparent similarity.
Although SCF also leads to vortex columns induced under the Coriolis effect, the transition typical in shear applied between spheres that rotate differently is  initiated by vortices generated in a staggered manner about the equatorial plane.
This may be attributed to the presence of meridional circulations due to differential rotation, which is more influential in the case of a larger gap between spherical boundaries.
Thus, the symmetry that appears in the transitional stage is distinct from vortex columns induced by the rotation of the entire system\cite{Hol06}.
In SCF, rotating wave states with a few azimuthal wavenumbers are robust and sustained as temporally quasi-periodic states under an external axial magnetic field\cite{Gar18,Gar20}.

\IND{}
Here, we examine the robustness of the rotational wave state in a wide-gap SCF against unstable thermal stratification.
Through the current study, we also hope to apply  the authenticity of the state in  SCF  (confirmed both numerically and experimentally) to our uncertain knowledge of {  \it spherical B\'enard convection} (SBC) under the radial gravitational field, which has only been investigated numerically.
The second purpose was to confirm the potential of residual shear stress between the boundaries of the SCF under thermal convection.
Examining the transition of laminar flow to rotating wave states perturbed by the thermal factor reconfirmed a phase diagram, which was obtained in Ref.\cite{Ina19}, in the extended control parameter space.

\IND{}
The remainder of this paper is organized as follows.
We first formulate our SCF and SBC hybrid system.
We describe the numerical method used for time development followed by describing the homotopy continuation to specify the rotating wave state in the hybrid system.
After commenting on the measures of symmetry, we explain the path in the parameter space.
We then give details of the numerically obtained states.
We then discuss the extent to which non-axisymmetry and rotation are destroyed around the polar axis at the continuation limit.

\SEC{Numerical Method}
\IND{}
We consider a radioactive Boussinesq fluid confined in a spherical shell with a thickness of $2\Delta \tilde{r}$ and mean radius $\tilde{r}_0$. The fluid was subjected to shear applied by the differential rotation of the boundaries. We assume that the mass and internal heat sources are distributed homogeneously inside the entire sphere as well as within the shell.
The heat radiation at the outer surface of the shell may generate a nonequilibrium state in the shell with a temperature difference of $2\Delta \tilde{T}$ between its inner and outer surfaces. The length, time, and temperature may be nondimensionalized by $\Delta \tilde{r}$, diffusion time $\Delta \tilde{r}^2/\tilde{\nu}$, and $\Delta \tilde{T}$, respectively, where $\tilde{\nu}$ is the kinematic viscosity.
The nondimensionalized equations \cite{Ina19} that govern the velocity and temperature fields, $\bfv{u}$ and $\varTheta$, are
\[
  \frac{{\rm D}\bfv{u}}{{\rm D}t}=-\bfv{\nabla}p+\bfv{\nabla}^2\bfv{u}+\Grs \varTheta r \bfv{e}_r \ \ ,
  \frac{{\rm D}\Theta}{{\rm D}t}=\frac{1}{\Prt}\Bigl(\bfv{\nabla}^2\varTheta + 3\frac{1-\eta}{1+\eta} \Bigr) \ \ .
\]
The third term on the right-hand side of the first equation, the gravity acceleration proportional to the radius, is attributed to the homogeneity of the mass distribution, while the first term on the right-hand side of the second equation is the homogeneity of the heat source in the inner sphere as well as the shell.
The no-slip boundary and isothermal conditions are adopted, that is, $\bfv{u}=(r_0-1) \Omega_{\rm in}\bfv{e}_\phi$ and $\Theta=1$ at the inner surface, and $\bfv{u}=0$ and $\Theta=-1$ at the outer surface of the shell, where $r_0=\tilde{r}_0/\Delta \tilde{r}$.
The system is uniquely determined by four dimensionless parameters: Prandtl number $\Prt$, Grashof number $\Grs$ that is proportional to the applied radial temperature difference, radius ratio $\ds \eta=\frac{r_0-1}{r_0+1}$, and Reynolds number $\ds \Rey=\frac{(r_0-1)^2\Omega_{\rm in}}{\nu}$.
In this study, we focused on the cases of $\Prt=1$ and $\eta=1/2$.
  
\IND{}
For the divergence-free constraint on $\bfv{u}$, we invoke the conventional toroidal and poloidal decompositions for the radial direction:
\[ \bfv{u}=\bfv{u}_{\rm S}+\bfv{\nabla}\times\bigl(-\bfv{r}\varPsi+\bfv{\nabla}\times(\bfv{r}\varPhi)\bigr)\ \ ,\]
where $\bfv{u}_{\rm S}=-\bfv{\nabla}\times\bigl(\bfv{r}\varPsi_{\rm S}\bigr)$  is Stokes flow, $\bfv{u}_{\rm S}=u_{\rm S}(r,\theta)\bfv{e}_\phi$, which contains neither radial nor polar  components and satisfies the aforementioned no-slip boundary conditions.
Note that the function $\varPsi_\phi$ is determined only by the value of $\Rey$; thus, the system is implicitly dependent on $\Rey$ via the $\bfv{u}_{\rm S}$ introduced into the governing equation.
We spatially expanded the scalar fields $\varPhi$, $\varPsi$, and $\varTheta$ in terms of Chebyshev polynomials and spherical harmonics with the aid of open numerical libraries\cite{Sch13,Fri05}.
For instance, $\varPhi(r,\theta,\phi)=\sum_{l,m,n}\varPhi_{l,m,n}Y_l^m(\theta,\phi)T_n(y)$ using the relative radius $y=[-1,1]$, where $r=r_0 + y$.

\IND{}
By adapting the second-order Adams-Bashforth method accompanied by the Crank-Nicolson method for temporal discretization and the Gauss-Lobatto collocation method for the evaluation of nonlinear terms, we can convert the dimensionless governing equation to an equivalent inhomogeneous Helmholtz equation with at first order for $\varPsi$ and $\varTheta$ and second order for $\varPhi$.
These Helmholtz equations are equivalent to a set of linear algebraic equations for the expansion coefficients $\varPsi_{l,m,n}$, $\varPhi_{l,m,n}$, and $\varTheta_{l,m,n}$, which are solved using the LAPACK libraries\cite{And99}.
Unless indicated, we used the truncation levels $(l_{\rm max},m_{\rm max},n_{\rm max})=(30,30,32)$.
According to Ref.\cite{Wul99,Hol06,Nak02,Jun00}, for the purely spherical Couette flow with $\eta=1/2$, the basic axisymmetric state, namely ``0-vortex'' state, is known to lose stability to a non-axisymmetric rotating wave perturbation when the Grashof number is increased above a threshold value denoted by $\Rey_{\rm cr}$.
By starting the numerical integration with a Stokes flow with a small disturbance that was artificially generated by a series of random numbers, we also obtained nonequilibrium and rotating wave states around the transitional Reynolds numbers for $\eta=1/2$.
The developed numerical code\cite{Ina19} was validated for $\Rey<600$ and $\Grs=0$ using quantitative comparison with previous experimental and numerical results.
In this study, our attention was restricted to the rotating wave state generated over the transitional Reynolds number for a purely spherical Couette flow with a wide gap.

\IND{}
{
  We demonstrated a purely spherical Couette flow experiment with $\eta=1/2$ and reproduced the rotating wave states over transitional Reynolds numbers.
  In the expreriment, the rotating wave states were actually illuminated as steady pattern of aluminium flakes on a laser light sheet projected either on the azimuthal or meridional planes.
  The angular velocity of the experimentally obtained state is in qualitative agreement with that of the numerically obtained state.
  It should be emphasized that the rotating wave state with a constant angular velocity can be realized both numerically and experimentally.
}
Here, considering that this wave solution rotates with an unknown but constant angular velocity $\Omega_\phi$, we may evaluate $\frac{\partial}{\partial t}(\cdot)$ as $-\Omega_\phi \frac{\partial}{\partial \phi}(\cdot)$ in the set of linear algebraic equations for the constant expansion coefficients.
Additionally, the freedom of $\Omega_\phi$ for the solution can be specified by fixing the azimuthal phase of the solution.
By solving the deduced quadratic equations using the iterative Newton-Raphson method with the aid of LAPACK libraries, we can uniquely specify all the coefficients of the rotating solution and azimuthal angular velocity $\Omega_\phi$ simultaneously.

\IND{}
In the following section, we measure the degree of asymmetry for the states obtained numerically.
The degree of non-axisymmetry for the rotation axis and of the anti-symmetry for the equatorial plane may be measured by the following $L_2$ norm normalized by the corresponding quantity of Stokes flow:
\[L_2 := \frac{\int |\varPsi_{\rm AM,\rm 3D}|^2 dv}{\int |\varPsi_{\rm S}|^2 dv} \ \ ,\]
where $\ds \varPsi_{\rm AM,\rm 3D} := \sum_{l+m\in {\rm odd}, m\ne 0, n}  \varPsi_{l,m,n}Y_l^m(\theta,\phi)T_n(y)$ and we termed anti-mirroring as ``AM''.
The value of $L_2$ is zero for the axisymmetric state such as in Stokes flow, satisfying $\partial/\partial_\phi=0$.
Moreover, we measured the degree of rotation around the polar axis either by the flow rate across the meridional section $q_\phi$ or by the torque exerted on the outer sphere $T_z$.
After they are normalized by the corresponding quantity of the Stokes flow, these values are defined as 
\[q_{\phi}:=\frac{\int_{{\rm S}_1} d\bfv{s} \cdot \bfv{u} }{\int_{{\rm S}_1} d\bfv{s} \cdot \bfv{u}_{\rm S} } \ \ , \ \ T_z=\frac{\bfv{e}_z\cdot\int_{{\rm S}_2} \bfv{r}\times d\bfv{s}\cdot\bfv{\nabla}\bfv{u}}{\bfv{e}_z\cdot\int_{{\rm S}_2}  \bfv{r}\times d\bfv{s}\cdot\bfv{\nabla}\bfv{u}_{\rm S}} \ \ ,\]
where the surfaces for calculating the integral ${\rm S}_1$ and ${\rm S}_2$ are defined as $r=[r_0-1,r_0+1], \theta=[0,\pi], \phi=0$, and $r=r_0+1, \theta=[0,\pi], \phi=[0,2\pi]$, respectively.
Stokes flow should be a good approximation of purely spherical Couette flow, particularly at low Reynolds numbers.
The polar component of the torque acting on the outer sphere by the flow is analytically provided in Ref.\cite{Lan87}.

\SEC{Results}
\IND{}
As mentioned before, over the transitional Reynolds number in the system with $\Grs=0$ (SCF), the system prefers non-axisymmetric rotating waves compared to the basic axisymmetric state, which has been performed even in experiments\cite{Egb95}.
We refer to the rotating wave state with the wavenumber $m$  as the $m$-fold (spiral) state.
Here, note that the $m$-fold state satisfies the symmetries $\ds \varPhi(r,\theta,\phi+2\pi/m)=\varPhi(r,\theta,\phi)$ and $\varPsi(r,\theta,\phi+2\pi/m)=\varPsi(r,\theta,\phi)$.
The preferred state also satisfies shift-and-reflection symmetry with respect to the equatorial plane; that is, the fields in both hemispheres shift toward each other by half the wavelength, $\ds \varPhi(r,\pi-\theta,\phi+\pi/m)=\varPhi(r,\theta,\phi)$ and $\ds \varPsi(r,\pi-\theta,\phi+\pi/m)=-\varPsi(r,\theta,\phi)$.
At the limit of $\Grs=0$, the temperature field passively advected in the governing equation in the SCF inherits symmetry from $\varPhi$.
Thus, it is natural to assume that $\ds \varTheta(r,\theta,\phi+2\pi/m)=\varTheta(r,\theta,\phi)$ and $\ds \varTheta(r,\pi-\theta,\phi+\pi/m)=\varTheta(r,\theta,\phi)$ are satisfied in continuation of the $m$-fold state to the hybrid system with $\Grs\ne 0$.
In a previous study\cite{Got21}, we showed that the $3$-fold state bifurcates from the axisymmetric state at a slightly higher Reynolds number than the 4-fold state. We also showed that, with an increase in the Reynolds number, the formation of a basin of attraction of the $3$-fold state rapidly expands compared to that of the $4$-fold state.
In particular, the latter result should be emphasized, which explains how the latecomer, the $3$-fold state, tends to be  established experimentally as a stable state.
This bifurcation aspect  is expected even in the system $\Grs\approx 0$.

\begin{figure}[h]
  \centering
  \includegraphics[angle=0,width=0.98\columnwidth]{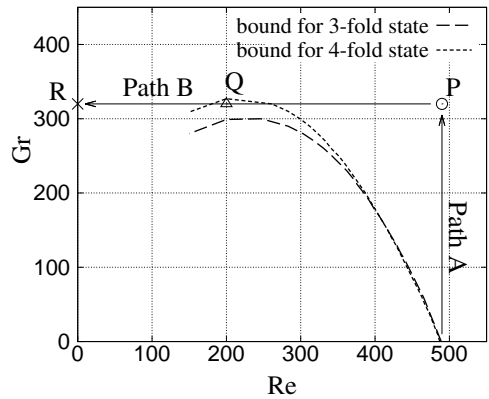}
  \caption{
    A square path starting at $(\Rey,\Grs)=(490,0)$ was used for the continuation of the 3 and 4-fold states.
    The path consists of two segments: (A) $\Rey=490$, $\Grs = 0\to 320$, and (B) $\Rey=490 \to 0$, $\Grs=320$, which has the three points P(490,320), Q(200,320), and R(0,320).
    The dashed and dotted curves indicate the neutral curves for the $m$-fold state $\Rey=\Rey_{\rm cr}^m(\Grs)$.
  }
  \label{map}
\end{figure}

\IND{}
Hereafter, we will extend the 4- and 3-fold states obtained at $\Grs=0$ towards the region of the $\Grs>0$ space.
The phase diagram in the space $(\Rey,\Grs)$ has been tentatively explored through a numerical  survey in Ref.\cite{Ina19}, although there was no guarantee whether the m-fold states are sustained in $\Grs>0$.
Here, the continuation of the 4- and 3-fold states is performed through a square path starting at $(\Rey,\Grs)=(490,0)$, as shown in Fig.\ref{map}.
The path consists of two segments: (A) $\Rey=490$, $\Grs = 0\to 320$, and (B) $\Rey=490 \to 0$, $\Grs=320$, which has the three representative points P(490,320), Q(200,320), and R(0,320).
In the present study, the neutral curves $\Rey=\Rey_{\rm cr}^m(\Grs)$ were calculated independently for the 4- and 3-fold states, respectively, which are indicated by the dashed and dotted curves in the figure.
The $m$-fold states bifurcate supercritically from the axisymmetric state on the neutral curves, and their infinitesimal nonaxisymmetric components exponentially decay for $\Rey<\Rey_{\rm cr}^m(\Grs)$ at a given $\Grs$.
A previous study\cite{Dum94,Jun00,Got21} showed that 4- and 3-fold states bifurcate from the axisymmetric state at
 $\Rey_{\rm cr}^4(0)=489$ 
 and $\Rey_{\rm cr}^3(0)=491$ 
 at SCF.
The present study shows that $\Rey=\Rey_{\rm cr}^m(\Grs)$ for both $m=3$ and $4$ monotonically decreased as $\Grs$ increased with an intersection at $(\Rey,\Grs)=(406,167)$.
The decrease in $\Rey=\Rey_{\rm cr}^m(\Grs)$ is intuitively consistent with the fact that  either the increase in $\Rey$  or the increase in $\Grs$ triggers the instability of the system.
Although $\Rey=\Rey_{\rm cr}^4(\Grs)$ passing once over segment A around $(261,320)$ reaches $\Grs>320$, $\Rey=\Rey_{\rm cr}^3(\Grs)$ exhibits a maximum value of approximately $200<\Rey<250$.
It is expected that, through a path that  circumvents the region\cite{Ina19}, the continuation of the $m$-fold state educed at the SCF towards the SBC limit is possible.

\begin{figure}[h]
  \centering
  \includegraphics[angle=0,width=0.98\columnwidth]{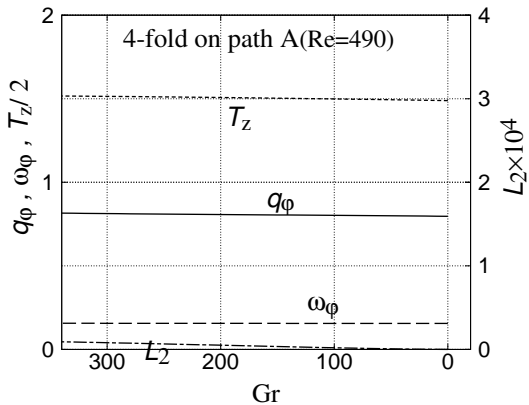}
  \caption{
    The continuous variation of the degrees of asymmetry to the 4-fold state with a continuous change of the parameter in segment A.
    The solid curve is the meridional circulation $q_\phi$, the dotted curve is half of the torque on the outer sphere $T_z/2$, the dashed curve is the normalized angular velocity $\omega_\phi$, and the dash-dotted curve represents the degree of nonaxisymmetric $L_2$.
  }
  \label{4-fold torque on path A}
\end{figure}

\IND{}
The non-axisymmetric component of the 4-fold state monotonically increases at the starting point of segment A, although it finally attenuates with a decrease in $\Rey$ on segment B to degenerate to the axisymmetric state around $(\Rey,\Grs)=(261,320)$ (see also Fig.\ref{4-fold torque on path A}).
However, the 3-fold state survives up to the end of segment B.
Thus, we only achieved the continuation of the 3-fold state from SCF to SBC, but not that of the 4-fold state.
The linear stability analysis at the purely thermal convective limit\cite{Cha61,Ita15} showed that the most dangerous mode in the conduction state consists of $Y_{l=4}^m$ and not $Y_{l=3}^m$ at $\Grs_{\rm cr}$ ($m=0,1,\cdots,l$) in the classical Rayleigh-B\'enard system in spherical geometry with $\eta=1/2$.
The replacement of the most dangerous mode from the 4-fold state at the SCF to the 3-fold state at the SBC on the path would seem to be somewhat puzzling.
However, it should be noted that the wavenumber of the most dangerous wave at the SCF is related to the azimuthal wave number $m$, but not to the polar wave number $l$ calculated by the linear stability analysis. 
The principal mode of the state achieved at point R based on the continuation of the 3-fold state is $Y_{l=4}^{m=3}$, as explained below.

\begin{figure}[h]
  \centering
  \includegraphics[angle=0,width=0.90\columnwidth]{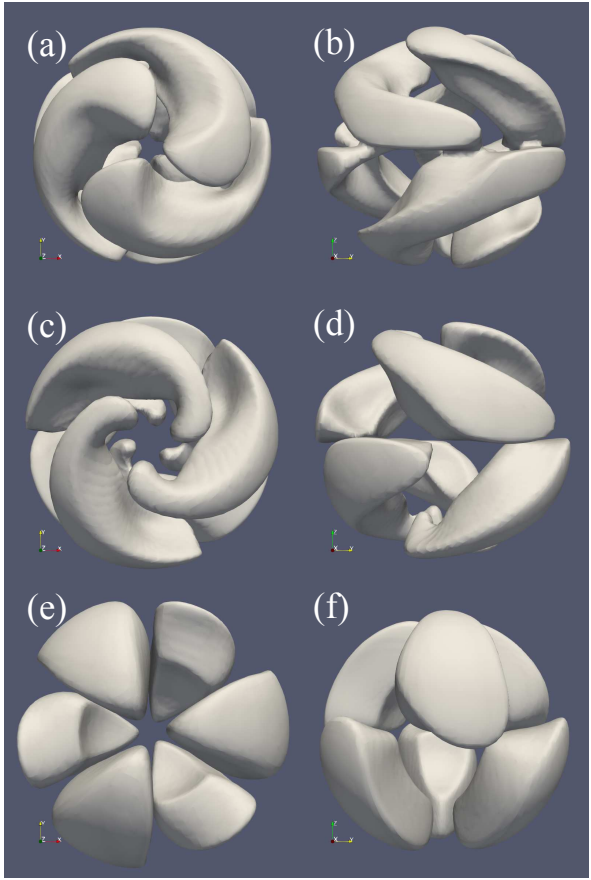}
  \caption{
    Polar and equatorial views of the 3-fold state at P (a,b), Q (c,d), and R (e,f) on segment B.
    The isosurface was $\varPhi_{\rm 3D}=C$, where $C$ was a tenth of the median of $\varPhi_{\rm 3D}$.
  }
  \label{photos}
\end{figure}

\IND{}
From the starting point of segment A to the middle of segment B, the 3-fold state sustains three spiral vortical arms alternately extending from the equator to the poles in each hemisphere, which are the same as those visualized in purely spherical Couette experiments at transitional Reynolds numbers\cite{Wul99}.
Fig.\ref{photos} shows the isosurface of $\varPhi_{\rm 3D}=C$ at point P,Q, and R on segment B, where $\varPhi_{\rm 3D}$ is the non-axisymmetric component of $\varPhi$.
Note that an isosurface qualitatively corresponds to a region of lift-up flow from the inner to the outer spheres for the non-axisymmetric component.
Close to point R from Q, the isosurface of the disturbance seems to gradually shift from a shear-related flow to a state originating at the heat convective flow.
As $\Rey$ vanishes, the amplitude of the disturbance in the polar zone diminishes and the 3-fold state continuously changes from a spiral pattern to a regularly convective pattern as six slightly skewed triangular pillars are positioned on the faces of a regular hexahedron, as  can be seen in Fig.\ref{photos}(f), which could be related to the mode $Y_{l=4}^{m=3}$.

\begin{figure}[h]
  \centering
  \vskip -1.5cm
  \includegraphics[angle=0,width=1.10\columnwidth]{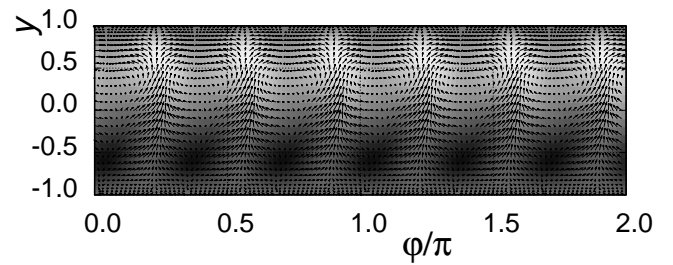}
  \vskip -1.5cm
  \caption{
    The 3-fold state that is achieved at $(\Rey,\Grs)=(0,320)$ by the continuation through the path.
    The disturbance component of $\varTheta$ and the flow field $(u_\phi,u_r)$ on the equatorial plane are described in the contour and vector fields, respectively.
  }
  \label{equatorial section 1} 
\end{figure}

\IND{}
At SBC ($\Rey=0$), the unstable modes that degenerate owing to spherical homogeneity emerge simultaneously from the static thermally conductive state over the critical Grashof number $\Grs_{\rm cr}$.
Owing to the nonlinear interactions  among these modes, highly symmetric steady states that are invariant under a set of transformations of point groups, such as axisymmetric or polyhedral patterns, may bifurcate directly from the static state\cite{Cha61,Bus75,Zeb83}.
For the top view (Fig.\ref{photos}(e)), it must be considered that the 3-fold state achieved at the point R additionally satisfies the mirror symmetry with respect to the $\phi=0$ plane, $(\varPhi,\varPsi,\varTheta)(r,\theta,\phi)=(\varPhi,-\varPsi,\varTheta)(r,\theta,-\phi)$.
However, the equatorial view (Fig.\ref{photos}(f)) suggests that the mirror symmetry is not practically satisfied by the 3-fold state obtained at R, where the Grashof number is significantly larger than $\Grs_{\rm cr}$.
The weak breaking of the mirror symmetry is associated with the nonzero angular momentum around the polar axis, as discussed later.
This was also clear in the equatorial section of the flow structure of the 3-fold state at R, as shown in Fig.\ref{equatorial section 1}.

\begin{figure}[h]
  \centering
  \vskip -1.5cm
  \includegraphics[angle=0,width=1.10\columnwidth]{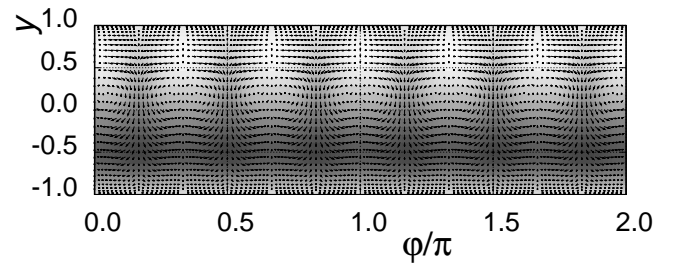}
  \vskip -1.5cm
  \caption{
    A saturated state was obtained by imposing mirror symmetry with respect to the $\phi=0$ plane on the 3-fold state obtained at R,
    as in Fig.\ref{equatorial section 1}.
  }
  \label{equatorial section 2} 
\end{figure}

\IND{}
The flow structure of the 3-fold state is a set of sinusoidal waves travelling in the azimuthal direction at the equatorial plane, which connects the spiral disturbance extending into each hemisphere from the poles.
For comparison, we examined the following computation.
Imposing the mirror symmetry with respect to the $\phi=0$ plane, time integration using the 3-fold state obtained at R as the initial condition saturates to another state similar to the 3-fold state at R.
The obtained state exactly  satisfies mirror symmetry with respect to the $\phi=0$ plane, which  was not  satisfied by the 3-fold state.
Figs.\ref{equatorial section 1} and \ref{equatorial section 2} show the disturbance component of $\varTheta$ and the flow field $(u_\phi,u_r)$ on the equatorial plane, which are described in contour and vector fields, respectively.
While Fig.\ref{equatorial section 1} is obtained from the $3$-fold state at R, Fig.\ref{equatorial section 2} is obtained from the additionally computed mirror-symmetric state at R.
A comparison of the figures indicates that the temperature and vector fields in the 3-fold state at R are distorted in the azimuthal direction and violate the mirror symmetry with respect to the $\phi=0$ plane.
The difference in magnitude in the temperature and vector fields was quantitatively nonnegligible; thus, the 3-fold state obtained at R is distinct from  the highly symmetric steady and polyhedral states invariant under a set of transformations of point groups.
This will be discussed in further detail in the next section.

\begin{figure}[h]
  \centering
  \includegraphics[angle=0,width=0.98\columnwidth]{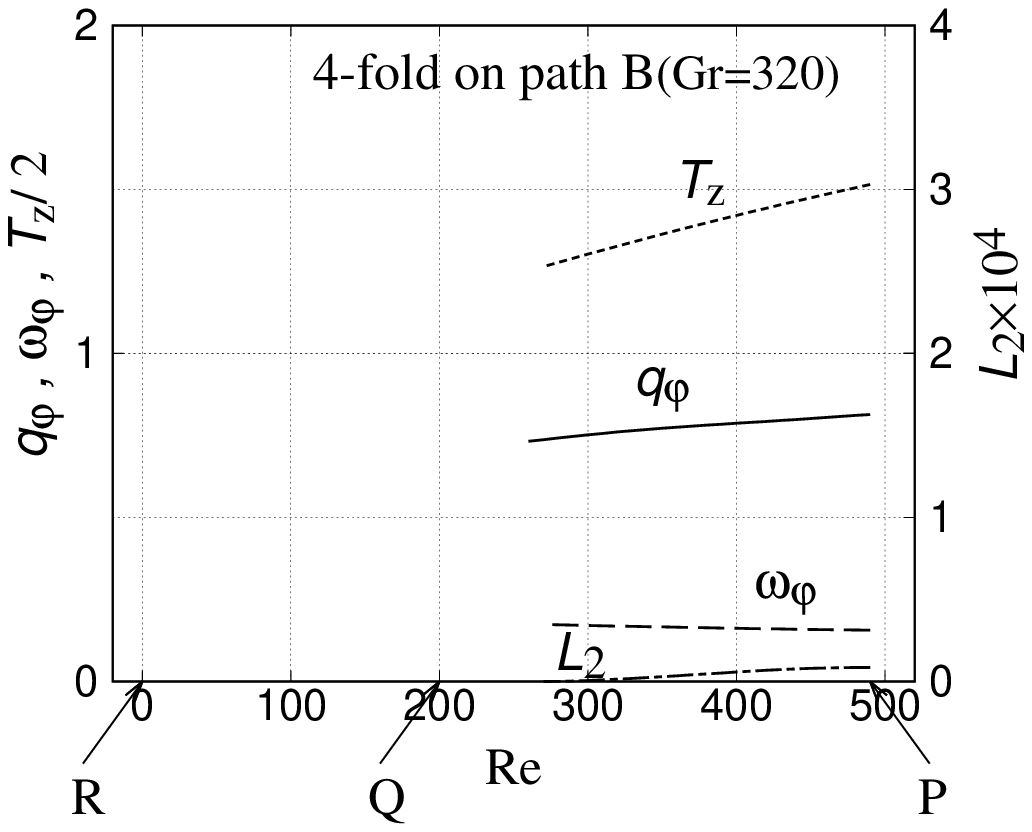}
  \caption{
    Continuous variation of the degrees of asymmetry to the 4-fold state with continuous parameter change in segment B, 
    as in Fig.\ref{4-fold torque on path A}.
  }
  \label{4-fold torque on path B}
\end{figure}
\begin{figure}[h]
  \centering
  \includegraphics[angle=0,width=0.98\columnwidth]{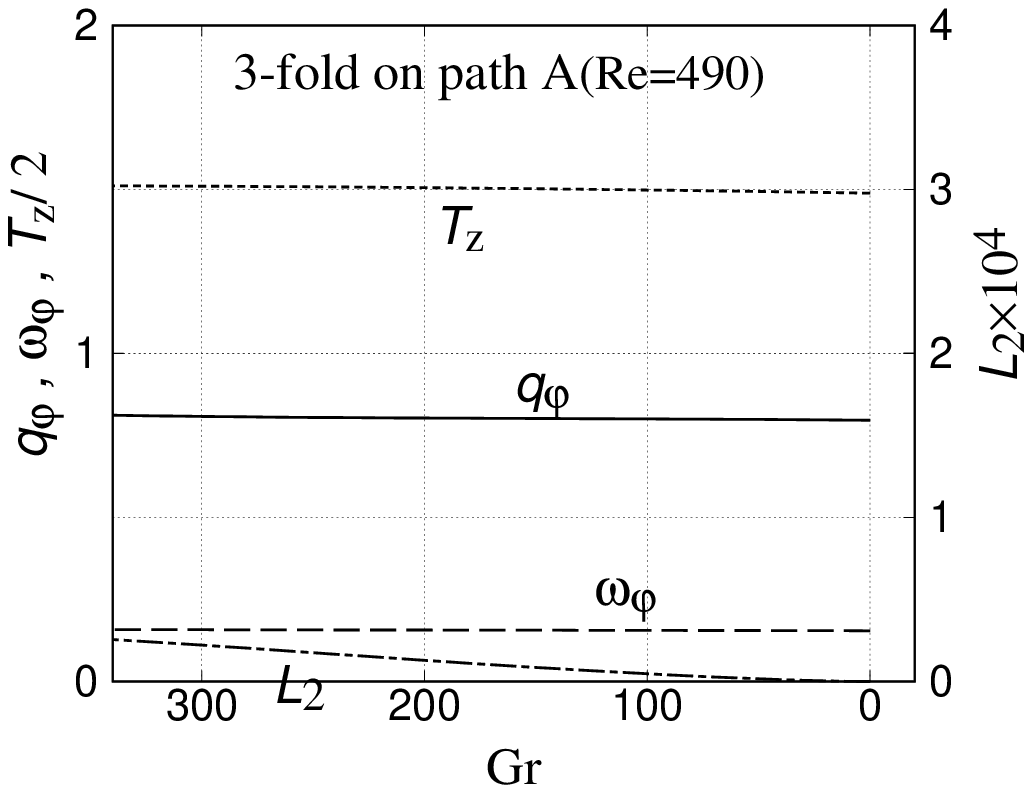}
  \caption{
    Continuous variation of the degrees of asymmetry to the 3-fold state with continuous parameter change in segment A,
    as win Fig.\ref{4-fold torque on path A}.
  }
  \label{torque on path A}
\end{figure}
\begin{figure}[h]
  \centering
  \includegraphics[angle=0,width=0.98\columnwidth]{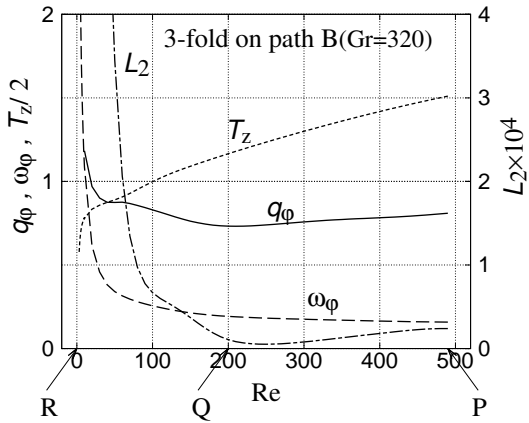}
  \caption{
    Continuous variation of the degrees of asymmetry to the 3-fold state with continuous parameter change in segment B,
    as in Fig.\ref{4-fold torque on path A}.
  }
  \label{torque on path B}
\end{figure}

\SEC{Discussion}
\IND{}
Figs.\ref{4-fold torque on path A} and \ref{4-fold torque on path B} show the continuous variation of the degrees of asymmetry with respect to the 4-fold state with continuous parameter change for segments A ($\Grs:0\to 320$) and B ($\Rey:490\to 0$).
Here, as an index of the degree of rotation with respect to the polar direction, we plotted the normalized angular phase velocity $\omega_\phi=\Omega_\phi/\Omega_{\rm in}$, $q_\phi$, and $T_z$.
Note that all the indices in both figures change continuously.
It does not appear that the branches of 4-fold states have discontinuous nodes on the path.
As shown in the figures, with an increase in $\Grs$ from $0$ to $320$ at $\Rey=490$ in segment A, the index for the degree of nonaxisymmetry, $L_2$, increases monotonically, whereas $L_2$ decreases monotonically until the degeneration to the axisymmetric state, with a decrease in $\Grs$ from $490$ to $261$ in segment B.
Because the system is rotated as far as $\Rey>0$, it is natural that the other indices describing asymmetry are finite throughout the path.

\IND{}
Figs.\ref{torque on path A} and \ref{torque on path B} show the continuous variation of the degrees of asymmetry for the 3-fold state with continuous parameter changes in segments A and B.
The branches of 3-fold states also have  no discontinuous node on the path.
As shown in Fig.\ref{torque on path A}, with an increase of $\Grs$ from $0$ to $320$ at $\Rey=490$ on segment A, the index for the degree of nonaxisymmetry, $L_2$, increases monotonically, while the degree of rotation with respect to the polar direction remains almost constant.
This suggests that the angular velocity of the state in the hybrid system with both $\Rey\ne 0$ and $\Grs\ne 0$ can be estimated only using $\Rey$.

\IND{}
In contrast, all the indices of the 3-fold state change continuously but largely in Fig.\ref{torque on path B}, where $\Rey$ decreases from $490$ to $0$ in segment B.
The index for the degree of nonaxisymmetry, $L_2$, monotonically increases after being annihilated around $\Rey=240$ with $\Rey$ decreasing in segment B.
The transient decrease in $L_2$ at $\Rey>240$ suggests that segment B passes through the neutral curve, $\Rey=\Rey_{\rm cr}(\Grs)$.
Note that the denominator of $L_2$ is proportional to the strength of the Stokes flow; that is, $\Rey$.
Thus, the divergence of $L_2$ to infinity at $\Rey\to 0$ in segment B implies that the numerator of $L_2$ converges to a nonzero value in the limit of $\Grs\to 0$.
This is not surprising because a variety of non-axisymmetric equilibrium states \cite{Cha61,Bus75,Zeb83} can be realized in the purely thermal convective state, $\Rey=0$.
%
%

\IND{}
In general, fluid mixing in the shell induced by the 3-fold state leads to some momentum exchange between the inner and outer spherical boundaries, so that the torque exerted on the outer sphere is larger than that of the Stokes flow; $T_z>1$ should be basically satisfied as  long as the truncation error is negligible.
The normalized torque $T_z$, which is maintained at approximately 3 on segment A converges to a finite value at $\Rey\to 0$ on segment B.
This convergence suggests that the torque exerted on the outer sphere vanished at $\Rey\to 0$.
Naturally, the torque exerted on the boundary from the fluid vanishes when the rotation of the system stops.
The symmetry is a coincidence between the solution and equation.

\IND{}
In contrast to the torque, the values of the normalized flow rate and angular velocity, $q_\phi$ and $\omega_\phi$, diverge at the limit of $\Rey\to 0$ in Fig.\ref{torque on path B}.
This is because the numerators of these values converged to finite values at $\Rey\to 0$.
Interestingly, the degree of rotation in the polar direction remains at nonzero values at the limit.
In other words, the orientation of the state did not change when the system became isotropic.
In general, we should clearly distinguish between the symmetry of the governing equation and the symmetry of the states realized under the equation.
In particular, with respect to thermal convection, there are many examples in which a realized state breaks the symmetry inherited from the system.
For instance, the oscillative flow reversals intermittently change the circulation direction in conventional Rayleigh-B\'enard system breaking the temporal homogeneity\cite{Kid97}.
A spiral state realized in either a spherical reaction-diffusion system\cite{Sig11} or Rayleigh-B\'enard system\cite{Ita15} breaks orientation homogeneity.

\IND{}
Associated with symmetry breaking, the natural convection in a vertical slot between parallel plates with a temperature difference\cite{Nag83} is suggestive of the present result.
For fluids with a low Prandtl number, such as liquid metal, the two-dimensional transverse vortex state bifurcated from the basic steady state principally becomes unstable against subharmonic perturbation at the second critical Grashof number.
If the system is restricted to a harmonic domain where the subharmonic perturbation cannot manifest itself, the transverse vortex state loses its stability to the three-dimensional traveling wave solution at the second critical Grashof number.
Surprisingly, the bifurcated three-dimensional solution is the wave solution traveling not in the vertical direction but the spanwise direction.
The solution has a finite spanwise momentum but exerts no friction on the plates\cite{NagIta03}.
Unfortunately, this phenomenon cannot be confirmed because it is realizable in a virtual slot with periodic boundaries in both the vertical and spanwise directions.
However, it would have been noticed if it were confirmed in the spherical shell periodically in both the polar and azimuthal directions.

\IND{}
The thick atmosphere of Venus is known to rotate at speeds up to 60 times that of slow planetary spin\cite{Hor20}.
One would think that superrotation cannot be sustained in nature unless the atmospheric angular momentum around its spin axis is supplied by axially asymmetric waves or turbulence.
Otherwise, the diffusion between the atmosphere and the surface and the mixing due to circulation in the meridional plane induced by the pole-equator temperature difference should attenuate the angular momentum.
The present result is fluid-dynamical  evidence, implying that such a super- or sub-rotation may be sustained without some exchange of angular momentum via torque between the fluid and boundary.

\IND{}
In the spherical thermal convection system, the multiple unstable modes originally degenerated under homogeneity are revealed simultaneously  at a critical Grashof number, estimating the ratio of the buoyancy to viscous force.
Various types of highly symmetric steady states are invariant under a set of transformations of point groups, such as axisymmetric or polyhedral patterns, and bifurcate from the static state via nonlinear interactions between these modes; thus, the isotropy inherited in the system is first broken\cite{Cha61,Bus75,Zeb83}.
An arbitrary incompressible field may be decomposed into toroidal and poloidal components in the radial direction.
The velocity fields realizable over the threshold consist only of the poloidal field, because of the absence of an energy source term that sustains the toroidal component\cite{Ita15}.
This context is analogous to the question of how super-rotation is sustained on Venus and  to the deduction that the velocity field bifurcated sequentially with a quasistatic increase in the Grashof number never leads to any net  angular momentum.
The present study provided a counterexample against such an intuitive deduction, and showed that spherical thermal convection  is a realizable system that can sustain  a steady state with no momentum exchange between the fluid and boundaries, as presented in Ref.\cite{NagIta03}.

\SEC{Summary}
\IND{}
Here, we revisit the conventional SCF, where a differential rotation of the spherical boundaries generates a nonequilibrium state of an incompressible Newtonian fluid in the spherical shell.
From previous experimental and numerical studies on SCFs, it is known that the first transition in a relatively wide-gap SCF is initiated by the rotating wave states accompanied by a change in the net angular momentum around the rotation axis.
By introducing the Boussinesq effect and temperature difference between the boundaries to the equations of SCF, the existence of the rotating wave state can be extended into the parameter space spanned by the Grashof number.
The main interest of the present study was the homotopy continuation of the rotating wave state realized experimentally towards the thermal convection limit in the parameter space.
Recalling a clear distinction between the symmetry of a governing equation and the symmetry of the states realized under the equation, we should recognize that there may be no basis to deny the mechanism that sustains a finite net angular momentum.
For example, it was verified in experiments\cite{Ben90} that the thermal convection system of binary fluid mixtures in an annular container with a large aspect ratio forms a localized traveling wave that breaks reflection symmetry.
In the present study, we obtained a 3-fold spiral state at the vanishing limit of the system’s rotation.
The state obtained at the limit was asymmetric to the azimuthal equation and we provided a numerically exact solution of a spherical Rayleigh-B\'enard system.
Further research is needed to analyze the bifurcation of the state at $\Grs=0$.

\acknowledgments
The authors would like to thank Mr. T. Inagaki for his pilot survey and Dr. H. Yamashita for his valuable comments on the draft.
We would also like to thank Editage (www.editage.com) for English language editing.
This work was supported in part by the Grant-in-Aid for Scientific Research(C), JSPS KAKENHI  Grant No.20K04294.
This work also benefited from interaction within RISE-2018 No.824022 ATM2BT of the European Union Horizon 2020-MSCA program, which includes Kansai University.

\bibliography{vns06}
\bibliographystyle{unsrt}

\end{document}